# Ultrafast Plasmonic Rotors for Electron Beams

Fatemeh Chahshouri[1,*], Nahid Talebi[1,2,*]

[1]*Institute of Experimental and Applied Physics, Kiel University, 24098 Kiel, Germany*
[2]*Kiel, Nano, Surface, and Interface Science − KiNSIS, Kiel University, 24098 Kiel, Germany*

E-Mail: talebi@physik.uni-kiel.de; chahshouri@physik.uni-kiel.de;

**Abstract:**

The interaction between free electrons and laser-induced near-fields provides a platform to study ultrafast processes and quantum phenomena while enabling precise manipulation of electron wavefunctions through linear and orbital momentum transfer. Here, by introducing phase offset between two orthogonally polarized laser pulses exciting a gold nanorod, we generate a rotating plasmonic nearfield dipole with clockwise and counterclockwise circulating orientations and investigate its interaction with a slow electron beam. Our findings reveal that the circulation direction of plasmonic fields plays a crucial role in modulating electron dynamics, enhancing coupling strength, and controlling recoil. Furthermore, synchronizing the interaction time of the electron beam with rotational dipolar plasmonic resonances results in significant transfer of angular momenta to the electron beams and deflects the electron wavepackets from their original trajectory. These findings highlight the potential of plasmon rotors for shaping electron wavepackets, offering promising applications in ultrafast microscopy, spectroscopy, and quantum information processing.

## 1 Introduction

Recent years, innovations in electron microscopy have revolutionized nanoscience, enabling atomic-scale insights into biological, chemical, and semiconductor materials [1]. Moreover, the integration of coherent electron beams with femtosecond laser pulses [2] has further advanced electron microscopy, enabling the exploration of quantum phenomena [3], ultrafast charge oscillations [4], [5], and nonequilibrium optical excitations [6], [7]. Electron-Driven Photon Sources (EDPHS [8], [9], [10], [11]) within electron microscopes additionally have pushed the field further, facilitating interferometry and time-resolved spectroscopy with femtosecond time resolution[12] without the need for external lasers. Such developments have unlocked new possibilities for studying plasmon resonances [13], [14], [15], exciton dynamics [12], and phonon behavior [6], thereby driving breakthroughs in electron holography [16], [17], phase retrieval [18], attosecond pulse trains [19], [20], and wave packet shaping [13], [21], [22], [23].

Shaped electron wave functions have been shown to allow precise control over quantum electrodynamic interactions, scattering processes, and Bremsstrahlung emission [24]. Furthermore, shaped electron beams can enhance x-ray generation [25], and enable the distinction between different quantum interference pathways [26]. This approach also leads to advancements in imaging resolution [27], [28], selective probing [29], [30], low-dose imaging [31], quantum computing [32], and enhancing data transmission [33]. While traditional methods, such as nanofabricated phase masks [34], [35], [36], [37], magnetic field [38], and phase plate [39] can manipulate electron wavepackets, they are limited in terms of speed, active controlling, and concomitant transverse and longitudinal phase modulation of the electron wavepackets.

Ultrafast electron microscopy (UTEM), where electron wavepackets are used to probe laser-induced excitations in matter has in addition led to coherent and spatiotemporal shaping of electron wave functions [40], [41], [42]. In principle coherent light can be used for modulation [43], [44] of both longitudinal [13], [45], [46], [47] and transverse wave functions [48], [49]. These interactions occur either in free space, through coupling via the ponderomotive force [50], [51], [52], [53], [54], [55], [56] of a light wave, or within the optical near-fields of nanostructures excited by laser pulses [50], [57], where the latter is known as Photon-Induced Near-field Electron Microscopy (PINEM) [4].

PINEM enables exploring the dynamics of near-field excitations by analyzing photon-electron longitudinal momentum exchange versus the delay between the electron wavepackt and light pulses [2]. In such interactions, the coupling strength [4], which governs energy exchange with electrons, can be enhanced by reducing mode volume, employing dielectric medium, increasing the longitudinal electric field, or extending interaction lengths [31]. Therefore, extended mode lifetimes to the picosecond range in systems like photonic crystals, or whispering gallery modes [14], [58] results in more quanta of energy exchange between laser and electron wavepackets, i.e., more PINEM peaks. However, resonant phase-matching, achieved by matching electron velocity with phase velocity of light in a prism [59] has also demonstrated the exchange of hundreds of photon quanta with single electrons over long distances. Slow electrons interacting with localized plasmonic fields [60], [61] can also enhance the coupling coefficient by increasing the effective interaction time [13], [62]. Furthermore, it has been shown that, beyond the near-field-mediated regime, the vector potential of freely propagating light waves in systems utilizing a single Hermite-Gaussian laser pulse [63], two-color Kapitza-Dirac effect [20] (stimulated Compton scattering), and optical beat waves [64] can also result in inelastic scattering of electron beams.

The electron temporal coherence relative to the light period determines the interaction regime for the modulation of the electron beam. With few-cycle THz pulses [65], microwaves [66] or radio waves [67] where temporal coherence of the electron wave packet is shorter than the light period ($\Delta t_e < t_{ph}$) the electron spatial distribution follows classical electron deflection. In contrast, for an electron temporal coherence longer than the light period for example in near-infrared light [40] interactions give rise to quantum dynamics, greatly influenced by the electromagnetic vector potential. Therefore, electrons experience diffraction when passing through periodic gratings formed by counter-propagating or obliquely illuminated laser beams [68], where for the latter even quantum interferences between sequential single-photon processes and direct two-photon processes influence the final shape of the electron wavepacket. Similarly, plasmonic Fabry–Perot cavities, formed by counter-propagating surface plasmon polaritons, can induce diffraction in the electron wave function [69]. Additionally, the Lorentz force generated by localized plasmons in gold nanorods [13], [21], [22], [62], acts as both a phase and amplitude grating, enabling elastic diffraction and inelastic energy transfer. It has been demonstrated that precise phase modulation can be achieved by controlling nanostructure configurations [22], topology [13], and size [13], along with the spatial profile of near-fields [21], [22].

In this work, we introduce plasmonic rotors as a novel platform for manipulating free-electron wave functions. Here, we investigate the interaction of a slow electron beam with plasmonic rotors and examine how the direction of circulating dipolar plasmons controls the longitudinal and transverse recoil of the electron wavepacket. The plasmonic rotors are generated by two orthogonal laser pulses with perpendicular polarizations and a $(\pm\frac{\pi}{2})$ phase offset, interacting with a gold nanorod. These rotors enable coherent and enhanced momentum transfer to electron wave packets by enhancing the interaction time and influencing the effective light frequency at the rest frame of the electron. By synchronizing the electron propagation with the plasmonic rotor, we demonstrate that its angular momentum and probability amplitude in both real and reciprocal space are significantly influenced by the direction of plasmon circulation. We further demonstrate that for clockwise (CKW) rotation, where the electron propagation direction aligns with the near-field oscillation, the coupling strength and consequently the momentum transfer is enhanced. In contrast, for counterclockwise (CCKW) rotation, the coupling strength decreases. This approach enables an additional degree of control over the ultrafast modulation of electron wave functions for transverse momentum as well as energy transfer, with applications in electron imaging [2], diffraction [70], [71], and spectroscopy[4]. For such applications, shaped electron beams, such as electron vortex beams [30], have the potential for enhancing electron microscopy technology, particularly in the study of magnetic and biological specimens [72].

## 2 Materials and methods

To investigate the interaction between a laser-induced near-field and a free electron wavepacket beyond the adiabatic approximation [73], we have developed a self-consistent numerical framework that simultaneously solves Maxwell and Schrödinger equations [50]. For simplicity, we consider here the electron-light interaction in a two-dimensional space (*x-y* plane), with the electron propagating along the *x*-direction. Therefore, we study the near-field effects at the apex of a nanorod confined to the *x-y* plane, assuming the nanorod has infinite height along the *z*-axis. Since the angle of incidence and the polarization of the light are both in the *x-y* plane, there is no dynamics along the z-axis and the system can be studied in two dimensions. The plasmonic near-field properties in this framework are computed at each time step using the finite-difference time-domain (FDTD) method, where the gold permittivity is modeled using a Drude model with two critical point functions [50]. Subsequently, the field components are interpolated from the Maxwell domain into the Schrödinger frame. The time evolution of the electron wavepacket, $(\psi(\vec{r}, t))$, in the vicinity of the laser-induced near-field is determined by solving the Schrödinger equation with the minimal-coupling Hamiltonian. After the interaction in the Schrödinger frame, the final electron wavepacket is analyzed to extract information on energy modulation and electron recoil. Finally, the inelastic scattering cross-section map is calculated using the final electron wavefunction $(\psi_f(x, y, t \to \infty))$, as [13]:

$$\sigma(E, \varphi) = \frac{d}{dEd\varphi}\langle\psi_f(x, y, t \to \infty)|\hat{H}|\psi_f(x, y, t \to \infty)\rangle = (m_0/\hbar^2)|\tilde{\psi}(E, \varphi; t \to \infty)|^2. \qquad (1)$$

Here, $m_0$ represents the electron mass, and $\hbar$ denotes the reduced Planck constant. The electron kinetic energy is defined as $E = \hbar^2(k_x^2 + k_y^2)/2m_0$, and the scattering angle is given by $\varphi = \tan^{-1}(k_x/k_y)$. $\tilde{\psi}$ is the wavepacket in the momentum space. Moving beyond the non-recoil approximation provides a more detailed perspective on the interaction, as it captures not only the longitudinal momentum distribution but also the amplitude modulation of the electron beam and its transverse momentum spread [13].

When an electron with an initial momentum $p_e = \hbar k_e$ interacts with a laser-induced near-field, it absorbs or emits $n$ quanta of photons from longitudinal component of the scattered filed. Consequently, its wavefunction evolves into a superposition of momentum states given by $p_e = \hbar \left( k_e + n \left( \omega_{ph}/v_e \right) \right)$. This process results in an energy comb, where the spacing between the peaks is determined by the photon energy $\hbar \omega_{ph}$. The probability amplitude ($|\psi_n(x,t)|^2$) for the exchange of $n$ quanta of energy between the electron wavepacket and the near-field light is obtained by expanding the wavefunction versus a Bessel series using as:

$$|\psi_n(x,t)|^2 \propto J_n^2(2|g|), \tag{2}$$

where $J_n$ is the $n$-th Bessel function of the first kind, and $g$ represents the coupling strength [4]:

$$g = (e/\hbar\omega_{ph}) \int_{-\infty}^{\infty} dx'\, \tilde{E}_x(x', y; \omega_{ph}) e^{-ix'\omega_{ph}/v_e} = (e/\hbar\omega_{ph}) \tilde{E}_x(k_x = \omega_{ph}/v_e, y; \omega_{ph}), \tag{3}$$

where $\tilde{E}_x$ denotes the Fourier-transformed-$x$ component of the scattered field, with respect to time and $x$-axis. The transverse field component, on the other hand, causes the lateral diffraction of the electron beam. The arrangement of diffraction orders at different energies is influenced by the electron velocity, the optical near-field momentum distributions, and the nanoparticle topology [13].

## 3 Results and discussion

In this work, we investigate the interaction between circulating plasmonic dipoles and a slow electron wavepacket. We demonstrate that applying two orthogonal laser pulses with a $\pm \frac{\pi}{2}$ phase offset introduces a phase relationship between the $x$ and $y$-polarized near-field dipoles. This phase difference generates time-dependent dipole moments within the nanorod, which are not merely linear oscillations but instead exhibit rotational behavior driven by the evolving temporal phase difference. As a result, when the scattered wave includes a rotating dipole aligned with the electron propagation direction, effective coupling with the electron wavefunction is achieved. Conversely, the dipole oscillates too fast in the rest frame of the electron and fails to facilitate a strong coupling.

We model the interaction of a Gaussian electron wavepacket with a kinetic energy of 1 keV, a longitudinal broadening of $W_L = 132$ nm, and a transverse broadening of $W_T = 15$ nm with two laser pulses at a central wavelength of $\lambda = 800$ nm and a temporal broadening of 21 fs (all values throughout the manuscript are full-wave at half maximum).

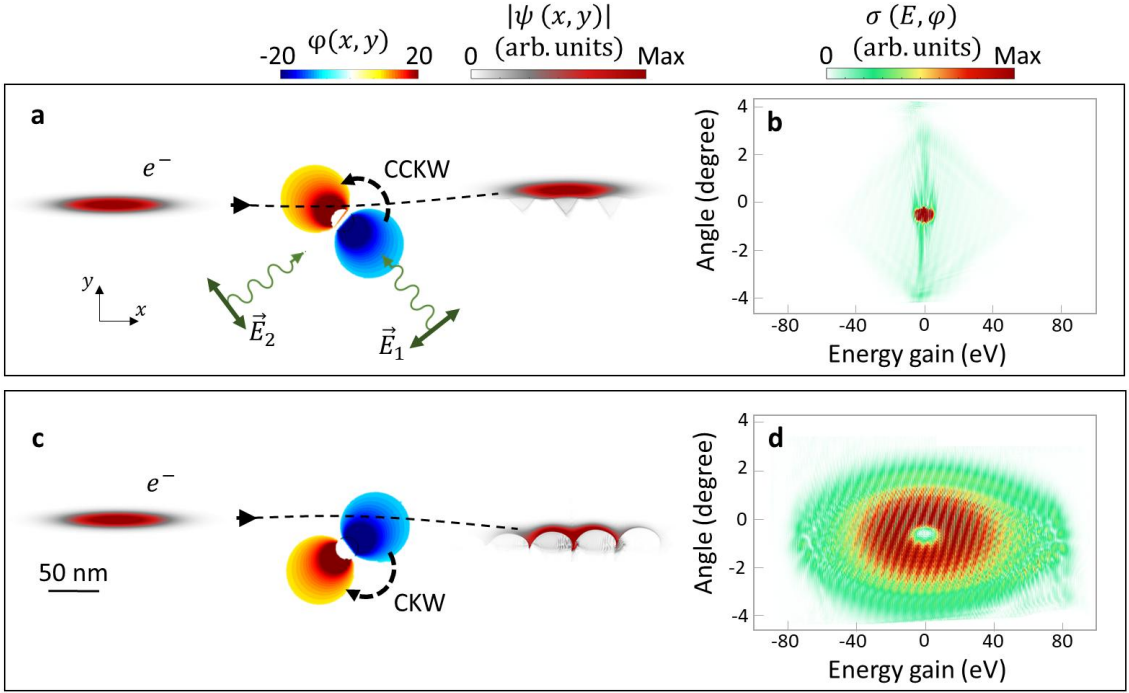

**Figure. 1. Electron beam shaping by a rotating localized plasmonic dipole.** The localized plasmon resonance is generated by two orthogonally polarized laser pulses, with a $\pm\frac{\pi}{2}$ phase offset between them, generating a CCKW or CKW plasmonic rotor depending on the phase offset. The angel-resolved inelastic scattering cross section of the electron wavepacket after the interaction with the (a, b) CCKW and (c, d) CKW dipolar modes of a gold nanorod. Panels (a, c) depict the modulation of the amplitude of the electron wavepacket in real space before and after interaction with the rotational near-field modes, while (b, d) illustrate the inelastic scattering cross-section following the interaction. The phase and direction of the optical near-field are represented by Re$\{\phi(Vx,y)\}$ and black curved arrow, where $\phi(x,y)$ denotes the scalar potential. The electron beam has an initial centre kinetic energy of 1000 eV, with longitudinal and transverse broadenings of 132 nm and 15 nm FWHM, respectively. The laser pulses feature a central wavelength of 800 nm, and FWHM temporal broadening of 21 fs, and a peak field amplitude of 2 GVm$^{-1}$. Dashed arrows indicate the trajectory of the electron along the *x* direction. The gold nanorod has a radius of 25 nm.

These initial parameters are selected to satisfy the synchronicity condition [13], $\lambda_{ph}v_{el} = 2R$ between the electron wavepacket propagation and effective dipolar mode oscillation, whereas keeping the interactions within the quantum regime, so that the longitudinal broadening is effectively longer than extend of the near-field [74]. Here, $\lambda_{ph}$ is the wavelength of the plasmonic resonances $v_{el}$ is electron velocity, and $R$ is the radius of gold nanorod ($R = 25$ nm). Figure 1 illustrates the electron modulation in both real and momentum space after the interaction with CCKW (a, b) and CKW (c, d) rotating near-field oscillations. The schematics on the left side of Figure 1 illustrate the amplitude modulation of the electron wavepacket in real space before and after interaction with the rotational near-field modes. The spatial distribution and the rotational direction of the optical near-field are represented by the scalar potential $\varphi(x,y)$ and the black curved arrow, respectively. As shown in Figure 1a, CCKW rotational field deflects the electron upward, increasing the impact parameter and reducing the efficiency of inelastic interactions. Conversely, the CKW dipolar oscillations deflects the electron downward (Figure 1c), decreasing the impact parameter. This reduction enhances the strength and efficiency of the interaction, resulting in more pronounced electron bunching and greater momentum transfer.

In addition, the number of oscillations of the projected that the electron observe along its propagation direction plays a crucial role in shaping the electron beam. The Lorentz force exerted by the near-field induces a wiggling motion in the electron wavepacket that further controls the interaction. The direction and dynamics of this wiggling motion plays a prominent role in the interaction strength and final extend of the wavepacket in the momentum space (see Supplementary Movie S1). For the case of CKW rotations, this force acts synchronously with the electron linear motion and leads to a unified transverse recoil across the energy distribution. Therefore, the rotational direction of wiggling motion controls the final electron modulation in momentum representation. In contrast, contrary-aligned field rotation relative to the electron propagation direction has a destructive effect. As a result, in the CCKW near-field case (Figure 1 b), momentum transfer spans diffraction angles up to $\varphi \approx \pm 3°$ within a small energy range of $-5\text{eV} \leq E \leq 5$ eV. In comparison, the CKW near-field (Figure 1d) induces stronger interactions. Consequently, the electron wavepacket experiences a transverse recoil spanning both positive and negative diffraction angles $(-4° \leq \varphi \leq +4)$, along with a broad longitudinal inelastic energy exchange within the range of $-80\text{eV} \leq E \leq 80$ eV. Studies of electron modulation with a single laser pulse have shown that the momentum transfer in the CKW system surpasses that of the *x*- and *y*-polarized light (Supplementary Figure 1). This enhanced transfer arises from the combined effects of both *x*- and *y*-polarized dipolar plasmons copropagating with the electron. In contrast, for CCKW field, the rotational restoring force against the electron propagation direction cancels out momentum exchange. Furthermore, repeating the simulations with a broader electron wavepacket yields similar results (Supplementary Figure 2).

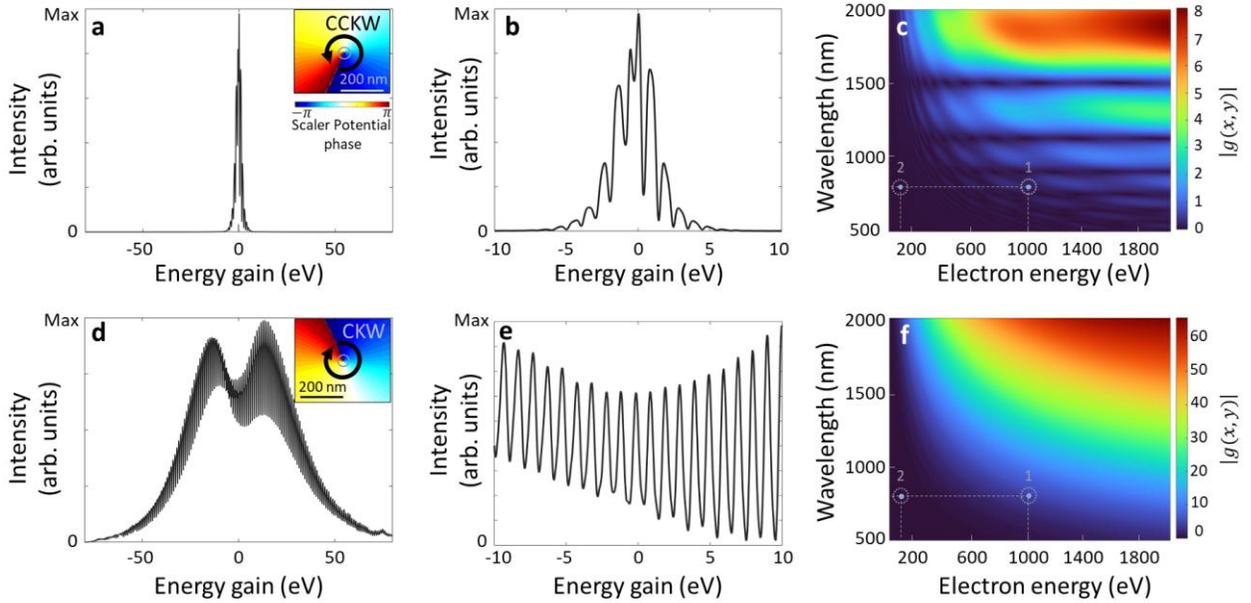

**Figure. 2. Impact of the near-field rotational direction on the energy transfer between light and free electron.** PINEM spectra at for an electron with the initial energy of 1 keV are shown after interaction with a (a, b) CCKW and (d, e) CKW rotational localized plasmonic dipoles. Panels (b) and (e) highlight the PINEM spectra in a narrow range near the zero-loss peak. The insets in (a) and (d) illustrate the simulated phase maps of the rotational scalar potential, derived from FDTD calculations. Panels (c) and (f) present the calculated coupling parameter *g* at the center of the electron beam for CCKW and CKW field oscillations, respectively.

Figure 2 better illustrates the role of the rotational direction of near-field in controlling the energy exchange between light and free electrons of the mentioned system. As demonstrated in insets of Figures 2a and d, representing the phase of the scalar potential, these circularly CCKW and CKW near-fields create a singularity at the center of the nanorod. These figures also present the PINEM spectra of the electron under time-varying near-fields, where several energy peaks are observed. Examining the energy spectrum near the zero-loss peak reveals distinct differences between the CCKW (Figure 2b) and CKW (Figure 2e) configurations. In the CCKW configuration, the small energy broadening ($\pm 5eV$) and irregular spacing between maxima indicate low coupling efficiency. Conversely, in the CKW system, precise phase-matching produces a fine spectral structure with well-defined spacing between each photon order. In the CKW configuration, dipolar field rotating along the electron beam propagation direction significantly amplifies the intensity of the PINEM energy spectra on both the gain and loss sides. In contrast, the counterclockwise design demonstrates the opposite effect, with reduced interaction strength and lower spectral intensity.

By calculating a map of the coupling coefficient $g$ (Eq. (3)) versus electron energies ranging from 20 to 2000 eV and photon wavelengths between 500 and 2000 nm, we observe the transition from the weak to strong coupling regime for rotational near-fields (see Figures 2 c and f). These maps illustrate how the interaction strength varies with the electron energy and the wavelength of the incident light, providing valuable insight into the dependence of the phase-matching criterion on the near-field properties. For instance, a slow electron beam with a kinetic energy of 100 eV interacting with a rotating dipole excited by an 800 nm laser wavelength (illustrated by point 2 in Figures 2 c and 2 f) lies in a weak interaction regime. In this case, the phase-matching condition for energy and momentum transfer cannot be fully achieved (see Supplementary Figure 3). Comparing the $g$-coefficient highlights how the rotation of the near-field breaks symmetry in the phase-matching criteria and enables selective energy transfer in either the gain or loss channels. In the CCKW (Figure 2c) system, the intensity of $g$ at the selected point (corresponding to the conditions in our simulation) is lower than in the CKW (Figure 2f) configuration. For the case of CCKW, dark lines appear in the map, marking regions of suppressed interaction. Moreover, for single *x*- or *y*- polarized laser-induced near-fields, the coupling strength is lower than that of the CKW configuration but higher than that of the CCKW configuration (see Supplementary Figure 1).

More intriguing electron shapes occurs when a diverging electron beam passes through the near-field region. In this scenario, we simulate a near-field area under the same initial conditions and electron energy as the previous system (Figures 1 and 2). However, the initial full-width at half-maximum (FWHM) of the longitudinal and transverse broadenings of the electron wavepacket are set to 15 nm and 132 nm, respectively. Figure. 3 illustrates the modulation of the diverging electron wavepacket under the influence of counterclockwise rotating (CCKW) (Figures 3a, b, c) and clockwise rotating (CWS) field (Figures 3d, e, f) configurations.

Within this framework, when the electron wavepacket enters the near-field region, its upper and lower parts experience a time-varying plasmonic field with a $\pi$ phase difference. This phenomenon is analogous to the Aharonov-Bohm effect; however, in this case, the scalar potential term of the Hamiltonian governs the interaction (assuming Coulomb gauge).

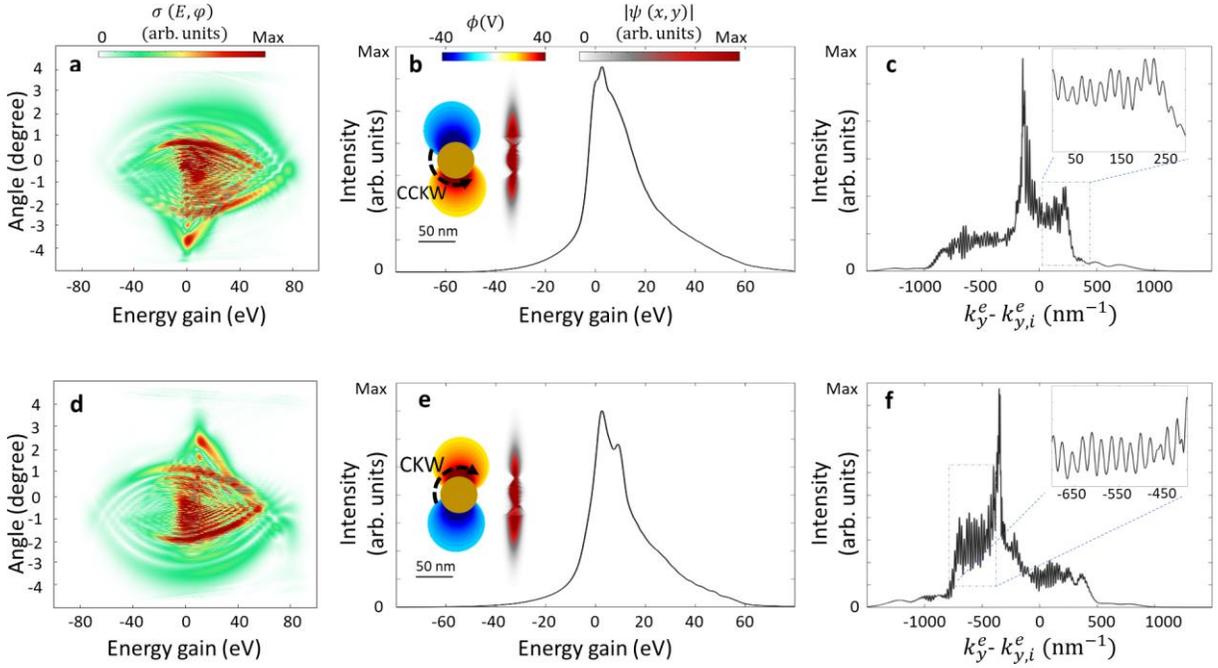

**Figure. 3. Deflection of a diverged electron beam influenced by near-field oscillation**. Electron modulation spectra after interaction with (a-c) right-handed and (d-f) left-handed rotating plasmons. (a, d) Inelastic scattering cross-section of the electron wavepacket after the interaction with the near-field. (b, e) PINEM spectrum, real-space distribution of the electron wavepacket, and a snapshot of the induced plasmonic near-field circulation orientation. (c, f) Transverse recoil of the electron beam integrated over the full energy range, with insets showing the magnified spectrum within a selected range. The electron beam is characterized by a kinetic energy of 1000 eV, with FWHM longitudinal and transverse broadenings of 15 nm and 132 nm. Where the laser pulses have a central wavelength of 800 nm, and a peak field amplitude of 2 GVm$^{-1}$.

The opposing sides of the electron wavepacket interact with near-field potentials of opposite signs, causing the wavepacket to split into two distinct paths. Finally, the interference between these two parts generates unique interference and diffraction patterns on the detector, as reflected in the final inelastic scattering cross-section map. As the diverging electron wavepacket passes through the center of the nanorod region, it undergoes four complete oscillations (the dynamics of this electron modulation are illustrated in Supplementary Figures 4 and 5). The overall phase accumulated by the electron over multiple light-field cycles, combined with the direction of field oscillation, determines the final momentum span of the wavepacket in both transverse and longitudinal directions. Consequently, the opposite oscillation directions in the clockwise and counterclockwise configurations result in vertically flipped momentum modulation maps (Figures 3a, and 3d).

The asymmetric force exerted by the oscillating fields causes significant transverse electron deflection after interaction with the near-field. The electron is deflected upward in the CCKW configuration and downward in the CKW configuration (see insets of Figures 3b and 3e, respectively). Moreover, since the electron stays in the interaction region for a short time and its longitudinal broadening is small, the inelastic momentum exchange is weak. Consequently, the final PINEM spectra for both configurations exhibit a broadband spectral feature, as shown in Figures. 3b and 3e. Along the transverse direction, the electron wavepacket experiences a significant Kapitza-Dirac-like diffraction as well. This near-field-mediated diffraction produces significant angular deflections in the transverse direction, surpassing those observed in the free-space Kapitza-Dirac effect ($2k_{ph}$), where inverse spectra for CCKW (Figure 3c), and CKW (Figure. 3, f) near-fields are observed.

Quasistatic approximations have been used elsewhere, when slow-electrons are used [62]. While this approximation provides a simplified framework for solving such systems, it falls short of offering a complete picture of electron recoil and its manipulation. As illustrated in Supplementary Figure 6, this method captures part of the main features described by the full Hamiltonian system.

The calculated coupling coefficient map reveals that electron-near-field interactions are significantly enhanced at higher electron velocities and lower near-field rotational speeds. The latter is evident from eq. (3), since the *g*-factor is inversely proportional to the light frequency, while the former is captured by the electron-photon interaction selection rule ($k_x = \omega_{ph}/v_e$). When phase-matching conditions between the electron and photons are satisfied, the electron recoil can be precisely manipulated, leading to strong modulation of the electron wavepacket.

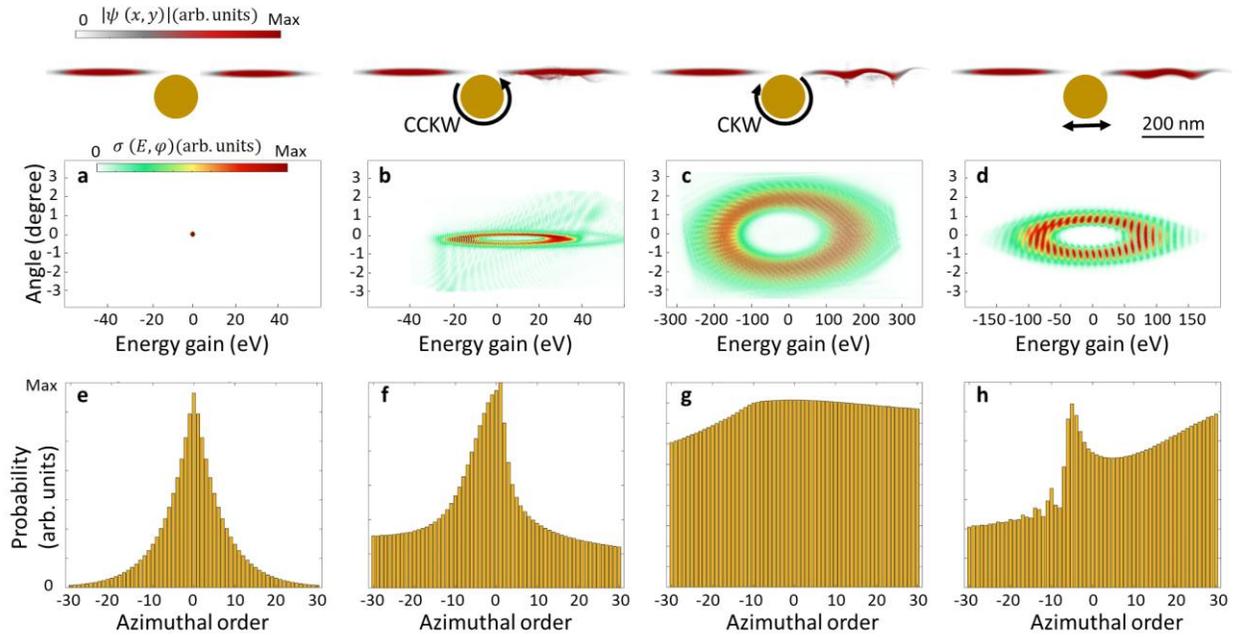

**Figure. 4. Probability amplitude distribution of the angular momentum transferred to the electron wavefunction by plasmonic near-field rotors.** A Gaussian electron wavepacket, with a kinetic energy of 1650 eV and transverse and longitudinal broadenings of 25 nm and 320 nm at FWHM, interacts with a plasmon generated by a nanorod with the radius of 80 nm. The inelastic scattering cross-section of the electron wavepacket is shown after propagation in (a) free space and interaction with (b) CCKW, (c) CKW, and (d) X-polarized plasmonic rotors. The upper row depicts the bunched electron profile after passing through the plasmonic near-field. Probability distribution of the angular momentum of the final electron wavefunction after propagating through (a) free space (no interaction), (b) CCKW, (c) CKW, and (d) x-oriented localized plasmonic dipolar fields. The laser pulse features a central wavelength of 2000 nm, an electric field amplitude of $E_0 = 1$ GVm$^{-1}$, and a temporal FWHM broadening of 53 fs, respectively.

This enhancement can be achieved by increasing both the nanorod radius and the wavelength of the incident light. Higher electron velocities and reduced near-field frequencies extend the effective interaction time between the copropagating field and the electron, enabling efficient electron-near-field coupling. Such coupling, critical for observing higher photon orders, is achieved under carefully optimized conditions. To investigate this phenomenon, we analyzed the influence of optical near-fields near a gold nanorod with a radius of 80 nm. The electron wavepacket is characterized by an initial energy of 1650 eV, with transverse and longitudinal broadenings of 25 nm and 320 nm, respectively. The incident laser wavelength is set at 2000 nm, and initial conditions is designed to ensure synchronicity between the electron and dipolar oscillations in both time-varying and static localized near-fields.

Figure 4 illustrates four configurations of the interaction: (a) no near-field, (b) CCKW near-field, (c) CKW near-field, and (d) *x*-polarized near-field, highlighting their respective influences on the propagating electron wavepacket. After the interaction the electron beam amplitude bunches and its linear and angular momentum deviate from the characteristics of a simple Gaussian beam (Figure 4a). By carefully analyzing the angular momentum probability distribution, we observed that this intense interaction effectively imparts angular momentum to the electron wavepacket.

The inelastic scattering cross-section map reveals distinct differences in energy gain/loss and angular distributions between CCKW (Figure 4b), CKW (Figure 4c), and *x*-polarized (Figure 4d) near-fields. For all interaction types, the free-electron wavepacket experiences a strong interaction regime, characterized by depletion of the ground state (zero-line peak) in the final modulation map. However, due to the horizontal broadening of the electron wavepacket, the influence of the *x*-polarized near-field is significant. For the *x*-polarized system, we observed a substantial energy transfer to the electron, spanning from -150 eV to 150 eV (supplementary Figure 7). Adjusting the phase matching by changing the rotational direction of the near-field for CCKW and CKW results in reduced or enhanced energy exchange and transverse diffraction, respectively. As a result, the CCKW field produces a symmetric and narrow energy gain/loss spectrum, indicative of low phase matching and coherent interaction with a small diffraction angle. In contrast, the CKW field causes a broader range of higher-order states, spanning within $-300 \text{ eV} < E < 300 \text{ eV}$ (supplementary Figure 7), with distinct peaks at elevated azimuthal orders.

To compute the angular momentum distribution of the final electron wavepacket, a Fourier expansion in terms of azimuthal angular orders is employed: $\psi(x, y, z) = \sum_m \psi_m(\rho, z) \exp(im\varphi)$, where $\rho = \sqrt{x^2 + y^2}$ is a radial component and $\varphi$ is the azimuthal angle. Then, the angular momentum probability distribution[73], restricted to the azimuthal order, is calculated as $P_m = \int_0^\rho d\rho \, dz \, |\psi_m(\rho, z)|^2$ to represent the azimuthal order distributions. However, we observe that the transformation of a single angular momentum order is not feasible; instead, the final electron wavepacket emerges as a complex superposition of multiple angular momentum orders (refer to Figures 4(e-h)). The CCKW field applies a symmetric rotational force, analogous to a central potential, resulting in coherent lower-order angular momentum transfer. This behavior mirrors classical Rutherford scattering patterns observed in small-angle deflections. In contrast, the CKW field generates an asymmetric, time-varying potential that scatters the electron wavepacket into higher-order angular momentum states, resembling high-energy scattering events with large-angle deflections. Similarly, panel (h) illustrates the azimuthal order distribution of the electron wavepacket under the influence of the *x*-polarized field. While linear polarization does not inherently carry angular momentum, a substantial angular momentum transfer is observed. This complexity arises because the *x*-polarized pulse creates a strongly oscillatory localized near-field that dynamically interacts with the electron wavepacket. The interplay between the electron motion and the non-uniform phase gradients in the near-field caused by plasmonic excitation and confinement generates these angular momentum distributions.

## 4 Conclusions

In conclusion, this study demonstrated the potential of plasmonic rotors as a powerful tool for manipulating free-electron wavepackets through controlled momentum transfer and energy modulation. By employing orthogonally polarized laser pulses with a phase offset, we excited circular dipolar near-fields in a gold nanorod and generated rotational plasmons with clockwise (CKW) and counterclockwise (CCKW) orientations. Our results reveal the intricate interplay between the direction of near-field rotation and the electron beam propagation, offering precise control over both linear and angular momentum exchange in the electron wavepacket. We showed that CKW fields significantly enhance energy transfer and electron recoil due to stronger phase matching and increased coupling efficiency, whereas CCKW fields exhibit narrower energy gain/loss distributions, indicative of reduced phase matching and weaker coupling. The unique characteristics of plasmonic rotors provide a versatile platform for advancing electron-beam shaping and harnessing coherent quantum interactions. These findings open new opportunities for integrating plasmonic rotors with other nanostructured materials to amplify coupling strength and expand the potential for high-resolution, low-energy electron microscopy. By extending these principles to more complex systems of rotational near-field excitations, this work lays a foundation for enhanced and active shaping of matter waves.

**Research funding**

This project has received funding from the European Research Council (ERC) under the European Union's Horizon 2020 research and innovation program, Grant Agreements No. 802130 (Kiel, NanoBeam) and No. 101017720 (EBEAM), as well as proof-of-concept Grant Agreements No. 101157312 (Kiel, UltraCoherentCL), as well as Momentum Grant of the Volkswagen Foundation.

**Conflict of interest**

Authors state no conflicts of interest.

**Author contributions**

N.T. initiated and supervised the project. F.C. conceived the idea and carried out simulations with N.T. F.C. and N.T analyzed the data and wrote the manuscript. Both authors have accepted responsibility for the entire content of this manuscript and approved its submission.

**Data availability**

The data that support the findings of this study are available from the corresponding author upon reasonable request.

# ORCID iDs

Nahid Talebi  https://orcid.org/0000-0002-3861-1005

Fatemeh Chahshouri https://orcid.org/0000-0001-5920-7805

# References

[1] A. Polman, M. Kociak, and F. J. García de Abajo, "Electron-beam spectroscopy for nanophotonics," *Nat Mater*, vol. 18, no. 11, pp. 1158–1171, 2019, doi: 10.1038/s41563-019-0409-1.

[2] B. Barwick, D. J. Flannigan, and A. H. Zewail, "Photon-induced near-field electron microscopy," *Nature*, vol. 462, no. 7275, pp. 902–906, 2009, doi: 10.1038/nature08662.

[3] A. Feist *et al.*, "Cavity-mediated electron-photon pairs," *Science (1979)*, vol. 377, no. 6607, pp. 777–780, Aug. 2022, doi: 10.1126/science.abo5037.

[4] S. T. Park, M. Lin, and A. H. Zewail, "Photon-induced near-field electron microscopy (PINEM): theoretical and experimental," *New J Phys*, vol. 12, no. 12, p. 123028, 2010, doi: 10.1088/1367-2630/12/12/123028.

[5] G. Hergert *et al.*, "Probing Transient Localized Electromagnetic Fields Using Low-Energy Point-Projection Electron Microscopy," *ACS Photonics*, vol. 8, no. 9, pp. 2573–2580, 2021, doi: 10.1021/acsphotonics.1c00775.

[6] Y. Kurman *et al.*, "Spatiotemporal imaging of 2D polariton wave packet dynamics using free electrons," *Science (1979)*, vol. 372, no. 6547, pp. 1181–1186, 2021, doi: 10.1126/science.abg9015.

[7] I. Madan *et al.*, "Charge Dynamics Electron Microscopy: Nanoscale Imaging of Femtosecond Plasma Dynamics," *ACS Nano*, vol. 17, no. 4, pp. 3657–3665, 2023, doi: 10.1021/acsnano.2c10482.

[8] N. van Nielen, M. Hentschel, N. Schilder, H. Giessen, A. Polman, and N. Talebi, "Electrons Generate Self-Complementary Broadband Vortex Light Beams Using Chiral Photon Sieves," *Nano Lett*, vol. 20, no. 8, pp. 5975–5981, 2020, doi: 10.1021/acs.nanolett.0c01964.

[9] N. Talebi, "Spectral Interferometry with Electron Microscopes," *Sci Rep*, vol. 6, no. 1, p. 33874, 2016, doi: 10.1038/srep33874.

[10] N. Talebi *et al.*, "Merging transformation optics with electron-driven photon sources," *Nat Commun*, vol. 10, no. 1, p. 599, 2019, doi: 10.1038/s41467-019-08488-4.

[11] J. Christopher, M. Taleb, A. Maity, M. Hentschel, H. Giessen, and N. Talebi, "Electron-driven photon sources for correlative electron-photon spectroscopy with electron microscopes," *Nanophotonics*, vol. 9, no. 15, pp. 4381–4406, 2020, doi: 10.1515/nanoph-2020-0263.

[12] M. Taleb, M. Hentschel, K. Rossnagel, H. Giessen, and N. Talebi, "Phase-locked photon–electron interaction without a laser," *Nat Phys*, vol. 19, no. 6, pp. 869–876, 2023, doi: 10.1038/s41567-023-01954-3.

[13] N. Talebi, "Strong Interaction of Slow Electrons with Near-Field Light Visited from First Principles," *Phys Rev Lett*, vol. 125, no. 8, p. 080401, 2020, doi: 10.1103/PhysRevLett.125.080401.

[14] K. Wang *et al.*, "Coherent interaction between free electrons and a photonic cavity," *Nature*, vol. 582, no. 7810, pp. 50–54, 2020, doi: 10.1038/s41586-020-2321-x.

[15] P. Das *et al.*, "Stimulated electron energy loss and gain in an electron microscope without a pulsed electron gun," *Ultramicroscopy*, vol. 203, pp. 44–51, 2019, doi: 10.1016/j.ultramic.2018.12.011.

[16] J. Vogelsang *et al.*, "Plasmonic-Nanofocusing-Based Electron Holography," *ACS Photonics*, vol. 5, no. 9, pp. 3584–3593, 2018, doi: 10.1021/acsphotonics.8b00418.

[17] I. Madan *et al.*, "Holographic imaging of electromagnetic fields via electron-light quantum interference," *Sci Adv*, vol. 5, no. 5, 2019, doi: 10.1126/sciadv.aav8358.

[18] T. Bucher *et al.*, "Free-electron Ramsey-type interferometry for enhanced amplitude and phase imaging of nearfields," *Sci Adv*, vol. 9, no. 51, 2023, doi: 10.1126/sciadv.adi5729.

[19] K. E. Priebe *et al.*, "Attosecond electron pulse trains and quantum state reconstruction in ultrafast transmission electron microscopy," *Nat Photonics*, vol. 11, no. 12, pp. 793–797, 2017, doi: 10.1038/s41566-017-0045-8.

[20] M. Kozák, N. Schönenberger, and P. Hommelhoff, "Ponderomotive Generation and Detection of Attosecond Free-Electron Pulse Trains," *Phys Rev Lett*, vol. 120, no. 10, p. 103203, 2018, doi: 10.1103/PhysRevLett.120.103203.

[21] F. Chahshouri and N. Talebi, "Numerical investigation of sequential phase-locked optical gating of free electrons," *Sci Rep*, vol. 13, no. 1, p. 18949, 2023, doi: 10.1038/s41598-023-45992-6.

[22] F. Chahshouri and N. Talebi, "Tailoring near-field-mediated photon electron interactions with light polarization," *New J Phys*, vol. 25, no. 1, p. 013033, 2023, doi: 10.1088/1367-2630/acb4b7.

[23] V. Di Giulio and F. J. García de Abajo, "Free-electron shaping using quantum light," *Optica*, vol. 7, no. 12, p. 1820, 2020, doi: 10.1364/OPTICA.404598.

[24] L. J. Wong *et al.*, "Control of quantum electrodynamical processes by shaping electron wavepackets," *Nat Commun*, vol. 12, no. 1, p. 1700, 2021, doi: 10.1038/s41467-021-21367-1.

[25] L. W. W. Wong and L. J. Wong, "Enhancing X-ray generation from twisted multilayer van der Waals materials by shaping electron wavepackets," *npj Nanophotonics*, vol. 1, no. 1, p. 41, 2024, doi: 10.1038/s44310-024-00043-4.

[26]   J. Lim, S. Kumar, Y. S. Ang, L. K. Ang, and L. J. Wong, "Quantum Interference between Fundamentally Different Processes Is Enabled by Shaped Input Wavefunctions," *Advanced Science*, vol. 10, no. 10, 2023, doi: 10.1002/advs.202205750.

[27]   T. Bucher *et al.*, "Coherently amplified ultrafast imaging using a free-electron interferometer," *Nat Photonics*, vol. 18, no. 8, pp. 809–815, 2024, doi: 10.1038/s41566-024-01451-w.

[28]   A. Béché, R. Juchtmans, and J. Verbeeck, "Efficient creation of electron vortex beams for high resolution STEM imaging," *Ultramicroscopy*, vol. 178, pp. 12–19, 2017, doi: 10.1016/j.ultramic.2016.05.006.

[29]   G. Guzzinati, A. Béché, H. Lourenço-Martins, J. Martin, M. Kociak, and J. Verbeeck, "Probing the symmetry of the potential of localized surface plasmon resonances with phase-shaped electron beams," *Nat Commun*, vol. 8, no. 1, p. 14999, 2017, doi: 10.1038/ncomms14999.

[30]   N. L. Streshkova, P. Koutenský, and M. Kozák, "Electron vortex beams for chirality probing at the nanoscale," *Phys Rev Appl*, vol. 22, no. 5, p. 054017, 2024, doi: 10.1103/PhysRevApplied.22.054017.

[31]   G. M. Vanacore, I. Madan, and F. Carbone, "Spatio-temporal shaping of a free-electron wave function via coherent light–electron interaction," *La Rivista del Nuovo Cimento*, vol. 43, no. 11, pp. 567–597, 2020, doi: 10.1007/s40766-020-00012-5.

[32]   O. Reinhardt, C. Mechel, M. Lynch, and I. Kaminer, "Free-Electron Qubits," *Ann Phys*, vol. 533, no. 2, p. 2000254, 2021, doi: 10.1002/andp.202000254.

[33]   R. Röpke, N. Kerker, and A. Stibor, "Data transmission by quantum matter wave modulation," *New J Phys*, vol. 23, no. 2, p. 023038, 2021, doi: 10.1088/1367-2630/abe15f.

[34]   H. Larocque *et al.*, "'Twisted' electrons," *Contemp Phys*, vol. 59, no. 2, pp. 126–144, 2018, doi: 10.1080/00107514.2017.1418046.

[35]   K. Y. Bliokh *et al.*, "Theory and applications of free-electron vortex states," *Phys Rep*, vol. 690, pp. 1–70, 2017, doi: 10.1016/j.physrep.2017.05.006.

[36]   J. Verbeeck, H. Tian, and A. Béché, "A new way of producing electron vortex probes for STEM," *Ultramicroscopy*, vol. 113, pp. 83–87, 2012, doi: 10.1016/j.ultramic.2011.10.008.

[37]   D. Roitman, R. Shiloh, P.-H. Lu, R. E. Dunin-Borkowski, and A. Arie, "Shaping of Electron Beams Using Sculpted Thin Films," *ACS Photonics*, vol. 8, no. 12, pp. 3394–3405, 2021, doi: 10.1021/acsphotonics.1c00951.

[38]   K. Y. Bliokh, P. Schattschneider, J. Verbeeck, and F. Nori, "Electron Vortex Beams in a Magnetic Field: A New Twist on Landau Levels and Aharonov-Bohm States," *Phys Rev X*, vol. 2, no. 4, p. 041011, 2012, doi: 10.1103/PhysRevX.2.041011.

[39]   P. Schattschneider, M. Stöger-Pollach, and J. Verbeeck, "Novel Vortex Generator and Mode Converter for Electron Beams," *Phys Rev Lett*, vol. 109, no. 8, p. 084801, 2012, doi: 10.1103/PhysRevLett.109.084801.

[40] Y. Morimoto and P. Baum, "Diffraction and microscopy with attosecond electron pulse trains," *Nat Phys*, vol. 14, no. 3, pp. 252–256, 2018, doi: 10.1038/s41567-017-0007-6.

[41] M. Kozák, T. Eckstein, N. Schönenberger, and P. Hommelhoff, "Inelastic ponderomotive scattering of electrons at a high-intensity optical travelling wave in vacuum," *Nat Phys*, vol. 14, no. 2, pp. 121–125, 2018, doi: 10.1038/nphys4282.

[42] G. M. Vanacore et al., "Attosecond coherent control of free-electron wave functions using semi-infinite light fields," *Nat Commun*, vol. 9, no. 1, p. 2694, 2018, doi: 10.1038/s41467-018-05021-x.

[43] G. M. Vanacore et al., "Ultrafast generation and control of an electron vortex beam via chiral plasmonic near fields," *Nat Mater*, vol. 18, no. 6, pp. 573–579, 2019, doi: 10.1038/s41563-019-0336-1.

[44] S. Tsesses et al., "Tunable photon-induced spatial modulation of free electrons," *Nat Mater*, vol. 22, no. 3, pp. 345–352, 2023, doi: 10.1038/s41563-022-01449-1.

[45] A. Feist, K. E. Echternkamp, J. Schauss, S. V. Yalunin, S. Schäfer, and C. Ropers, "Quantum coherent optical phase modulation in an ultrafast transmission electron microscope," *Nature*, vol. 521, no. 7551, pp. 200–203, 2015, doi: 10.1038/nature14463.

[46] J.-W. Henke et al., "Integrated photonics enables continuous-beam electron phase modulation," *Nature*, vol. 600, no. 7890, pp. 653–658, 2021, doi: 10.1038/s41586-021-04197-5.

[47] N. Talebi, "Schrödinger electrons interacting with optical gratings: quantum mechanical study of the inverse Smith–Purcell effect," *New J Phys*, vol. 18, no. 12, p. 123006, 2016, doi: 10.1088/1367-2630/18/12/123006.

[48] I. Madan et al., "Ultrafast Transverse Modulation of Free Electrons by Interaction with Shaped Optical Fields," *ACS Photonics*, vol. 9, no. 10, pp. 3215–3224, 2022, doi: 10.1021/acsphotonics.2c00850.

[49] M. C. Chirita Mihaila, P. Weber, M. Schneller, L. Grandits, S. Nimmrichter, and T. Juffmann, "Transverse Electron-Beam Shaping with Light," *Phys Rev X*, vol. 12, no. 3, p. 031043, 2022, doi: 10.1103/PhysRevX.12.031043.

[50] N. Talebi, *Near-Field-Mediated Photon–Electron Interactions*, vol. 228. in Springer Series in Optical Sciences, vol. 228. Cham: Springer International Publishing, 2019. doi: 10.1007/978-3-030-33816-9.

[51] D. L. Freimund, K. Aflatooni, and H. Batelaan, "Observation of the Kapitza–Dirac effect," *Nature*, vol. 413, no. 6852, pp. 142–143, 2001, doi: 10.1038/35093065.

[52] P. L. Kapitza and P. A. M. Dirac, "The reflection of electrons from standing light waves," *Mathematical Proceedings of the Cambridge Philosophical Society*, vol. 29, no. 2, pp. 297–300, 1933, doi: 10.1017/S0305004100011105.

[53] O. Reinhardt and I. Kaminer, "Theory of Shaping Electron Wavepackets with Light," *ACS Photonics*, vol. 7, no. 10, pp. 2859–2870, 2020, doi: 10.1021/acsphotonics.0c01133.

[54]	K. A. H. (Ton) van Leeuwen *et al.*, "Feasibility of a Pulsed Ponderomotive Phase Plate for Electron Beams," *New J Phys*, 2023, doi: 10.1088/1367-2630/acbc44.

[55]	S. Ebel and N. Talebi, "Structured free-space optical fields for transverse and longitudinal control of electron matter waves," 2024.

[56]	S. T. Kempers, I. J. M. van Elk, K. A. H. van Leeuwen, and O. J. Luiten, "Coherent electron phase-space manipulation by combined elastic and inelastic light-electron scattering," *New J Phys*, vol. 26, no. 9, p. 093026, 2024, doi: 10.1088/1367-2630/ad7631.

[57]	N. Talebi, "Interaction of electron beams with optical nanostructures and metamaterials: from coherent photon sources towards shaping the wave function," *Journal of Optics*, vol. 19, no. 10, p. 103001, 2017, doi: 10.1088/2040-8986/aa8041.

[58]	O. Kfir *et al.*, "Controlling free electrons with optical whispering-gallery modes," *Nature*, vol. 582, no. 7810, pp. 46–49, 2020, doi: 10.1038/s41586-020-2320-y.

[59]	R. Dahan *et al.*, "Resonant phase-matching between a light wave and a free-electron wavefunction," *Nat Phys*, vol. 16, no. 11, pp. 1123–1131, 2020, doi: 10.1038/s41567-020-01042-w.

[60]	R. Shiloh, T. Chlouba, and P. Hommelhoff, "Quantum-Coherent Light-Electron Interaction in a Scanning Electron Microscope," *Phys Rev Lett*, vol. 128, no. 23, p. 235301, 2022, doi: 10.1103/PhysRevLett.128.235301.

[61]	A. Wöste *et al.*, "Ultrafast Coupling of Optical Near Fields to Low-Energy Electrons Probed in a Point-Projection Microscope," *Nano Lett*, vol. 23, no. 12, pp. 5528–5534, 2023, doi: 10.1021/acs.nanolett.3c00738.

[62]	G. Hergert, A. Wöste, P. Groß, and C. Lienau, "Strong inelastic scattering of slow electrons by optical near fields of small nanostructures," *Journal of Physics B: Atomic, Molecular and Optical Physics*, vol. 54, no. 17, p. 174001, 2021, doi: 10.1088/1361-6455/ac2471.

[63]	S. Ebel and N. Talebi, "Inelastic electron scattering at a single-beam structured light wave," *Commun Phys*, vol. 6, no. 1, p. 179, 2023, doi: 10.1038/s42005-023-01300-2.

[64]	M. Kozák and T. Ostatnický, "Asynchronous Inelastic Scattering of Electrons at the Ponderomotive Potential of Optical Waves," *Phys Rev Lett*, vol. 129, no. 2, p. 024801, 2022, doi: 10.1103/PhysRevLett.129.024801.

[65]	C. Kealhofer, W. Schneider, D. Ehberger, A. Ryabov, F. Krausz, and P. Baum, "All-optical control and metrology of electron pulses," *Science (1979)*, vol. 352, no. 6284, pp. 429–433, 2016, doi: 10.1126/science.aae0003.

[66]	W. Verhoeven, J. F. M. van Rens, W. F. Toonen, E. R. Kieft, P. H. A. Mutsaers, and O. J. Luiten, "Time-of-flight electron energy loss spectroscopy by longitudinal phase space manipulation with microwave cavities," *Structural Dynamics*, vol. 5, no. 5, 2018, doi: 10.1063/1.5052217.

[67]  J. F. M. van Rens *et al.*, "Theory and particle tracking simulations of a resonant radiofrequency deflection cavity in TM 110 mode for ultrafast electron microscopy," *Ultramicroscopy*, vol. 184, pp. 77–89, 2018, doi: 10.1016/j.ultramic.2017.10.004.

[68]  N. Talebi and C. Lienau, "Interference between quantum paths in coherent Kapitza–Dirac effect," *New J Phys*, vol. 21, no. 9, p. 093016, 2019, doi: 10.1088/1367-2630/ab3ce3.

[69]  F. J. García de Abajo, B. Barwick, and F. Carbone, "Electron diffraction by plasmon waves," *Phys Rev B*, vol. 94, no. 4, p. 041404, 2016, doi: 10.1103/PhysRevB.94.041404.

[70]  P. Baum, "On the physics of ultrashort single-electron pulses for time-resolved microscopy and diffraction," *Chem Phys*, vol. 423, pp. 55–61, 2013, doi: 10.1016/j.chemphys.2013.06.012.

[71]  M. Eichberger *et al.*, "Femtosecond streaking of electron diffraction patterns to study structural dynamics in crystalline matter," *Appl Phys Lett*, vol. 102, no. 12, 2013, doi: 10.1063/1.4798518.

[72]  B. J. McMorran *et al.*, "Electron Vortex Beams with High Quanta of Orbital Angular Momentum," *Science (1979)*, vol. 331, no. 6014, pp. 192–195, 2011, doi: 10.1126/science.1198804.

[73]  N. Talebi, "Electron-light interactions beyond the adiabatic approximation: recoil engineering and spectral interferometry," *Adv Phys X*, vol. 3, no. 1, p. 1499438, 2018, doi: 10.1080/23746149.2018.1499438.

[74]  Y. Pan *et al.*, "Weak measurements and quantum-to-classical transitions in free electron–photon interactions," *Light Sci Appl*, vol. 12, no. 1, p. 267, 2023, doi: 10.1038/s41377-023-01292-2.

**Supplementary contents for**

**Ultrafast Plasmonic Rotors for Electron Beams**

Fatemeh Chahshouri[1,*], Nahid Talebi[1,2,*]

[1]*Institute of Experimental and Applied Physics, Kiel University, 24098 Kiel, Germany*
[2]*Kiel, Nano, Surface, and Interface Science − KiNSIS, Kiel University, 24098 Kiel, Germany*

E-Mail: talebi@physik.uni-kiel.de; chahshouri@physik.uni-kiel.de;

Content:

Supplementary Note 1: Shaping Electron Wave Packet with Linearly Polarized Near-Fields

Supplementary Note 2: Influence of the Electron Wavepacket Dimension

Supplementary Note 3: Effect of Electron Kinetic Energy on Its Final Modulation

Supplementary Note 4: Dynamics of Diverged Electron Beam Manipulation by Time-Varying Near-Field in CCWS System

Supplementary Note 5: Dynamics of the Interaction and the Electron Recoil in CWS System

Supplementary Note 6: Calculation of Electron-Photon Interaction with Quasi-Static Approximation

# Supplementary Note 1: Shaping Electron Wave Packet with Linearly Polarized Near-Fields

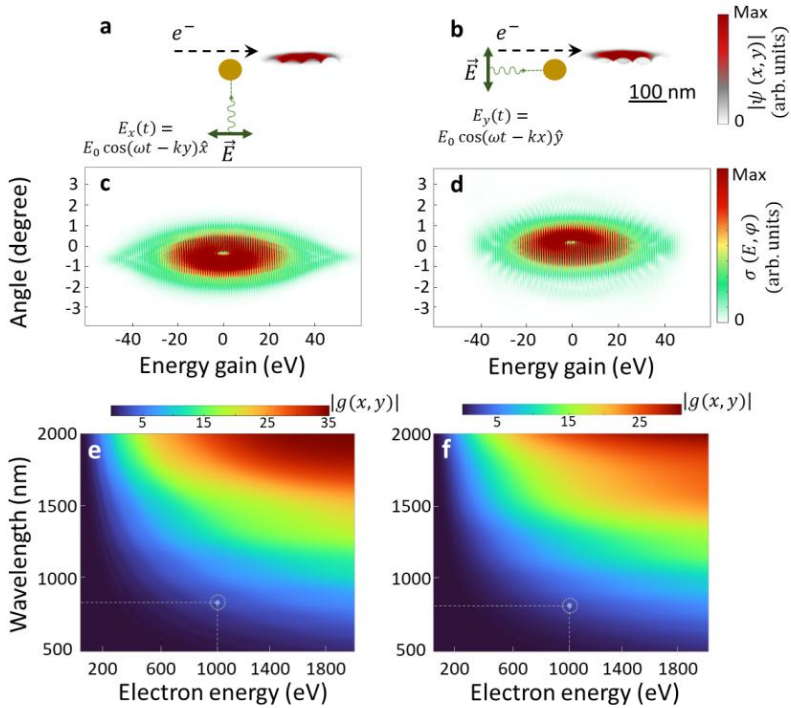

**Figure. S1. Modulation of the electron wavepacket after the interaction with linear-polarized dipolar plasmon oscillations in a gold nanorod**. (a) Amplitude of the electron wave function after the interaction with (a) x-polarized, (b) y-polarized near-fields. Inelastic scattering cross-section map corresponding to (c) x-polarized and (d) y-polarized excitation scenarios. The calculated coupling parameter *g* at the center of the electron beam for (e) x-polarized and y-polarized field oscillations, respectively.

To fully understand the impact of rotational plasmons on electron modulation, it is essential to compare them with the dynamics of the recoil exerted by a single linearly polarized gold nanorod. Figure S1 illustrates the electron response to the laser-induced plasmonic near-field of a single nanorod. In this analysis, the initial parameters of the electron and laser in the simulation domain are identical to those in the circular near-field scheme (Figures 1 and 2 of the main text), but each laser is applied separately. Figures S1a, c, and e depict the amplitude modulation of the electron wavepacket in real space, the momentum representation, and the coupling coefficient for the case of an *x*-polarized laser, respectively. Correspondingly, Figures. S1b, d, and f show the same responses for a *y*-polarized laser. Our numerical investigation reaffirms the critical role of the near-field excitation direction in manipulating the electron wave packet. Enhanced interactions near the nanostructure result in attosecond bunching of the electron wave packet. However, the projection of the scattered field along the electron trajectory emerges as a key factor in shaping the electron beam. For an *x*-polarized near-field interacting with a horizontally broadened electron wave packet, the scattered wave has a field component aligned with the electron beam direction, enabling stronger coupling. In contrast, when the electric field is perpendicular to the plane of the electron wave packet, as in the *y*-polarized system, the scattered wave does not facilitate the same level of coupling.

## Supplementary Note 2: Influence of the electron Wavepacket Longitudinal Broadening

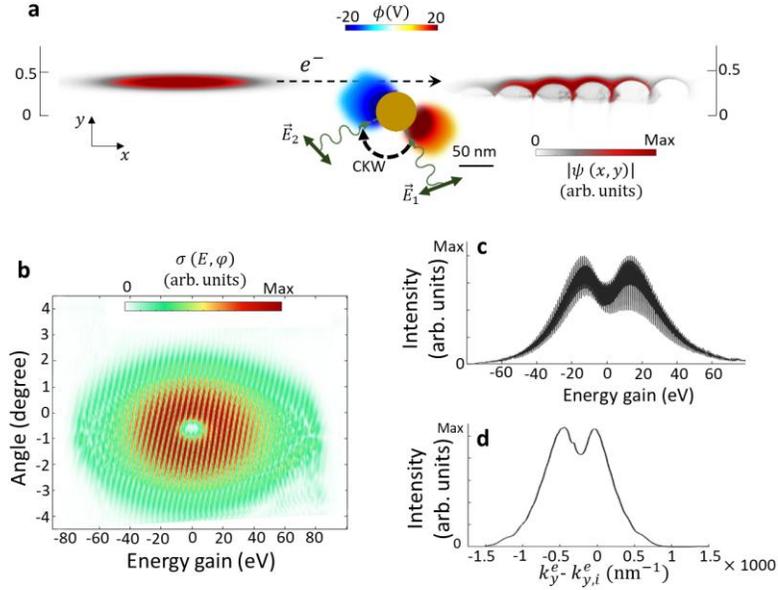

**Figure. S2. The effect of longitudinal broadening of the electron pulse on the final distribution of the electron wavepacket**. The final distribution of the electron wavepacket is shown in (a) real-space and (b) momentum-space after passing through a CKW rotational field, illustrated by the scalar potential map in panel (a). The electron wave packet has an initial kinetic energy of 1 keV and transverse and longitudinal broadenings of 15 nm and 200 nm, respectively. The lasers are characterized by an electric field amplitude of $E_0 = 2 \text{ GVm}^{-1}$, a wavelength of 800 nm, and an FWHM temporal broadening of 21 fs. Panel (c) displays the PINEM spectra after interaction with the system, while (d) shows the electron distribution along the transverse direction, integrated over the longitudinal momentum axis.

Here, we provide additional insights into the impact of the longitudinal broadening of the electron wave packet on its inelastic energy transfer (Figure S2c) and elastic diffraction recoil (Figure S2d). The simulation parameters are consistent with those used in Figures 1 and 2 of the main text, but with a longer electron pulse. Our simulations show that a more extensive spatial spread of the electron wave packet (Figure S2), corresponding to a longer effective interaction length with the clockwise rotational field (equivalent to 5 oscillation periods for 200 nm), aligns with the findings obtained for smaller wave packet broadenings. In both real-space (with the bunching shown in Figure S2a) and reciprocal-space (with an eye-like pattern of the momentum distribution in Figure S2b), the results remain consistent. The comparison emphasizes that the longitudinal phase modulation, arising from energy-momentum conservation, facilitates constructive interference paths. These paths, enabled by the longer electron pulse, sustain discrete and distinguishable momentum modulation, similar to those observed with shorter electron pulses. Furthermore, the transverse deflection of the electron in real-space becomes more pronounced for longer electron wave packet, indicating enhanced modulation due to the extended interaction time and a smaller impact parameter.

# Supplementary Note 3: Effect of Electron Kinetic Energy on Its Final Modulation

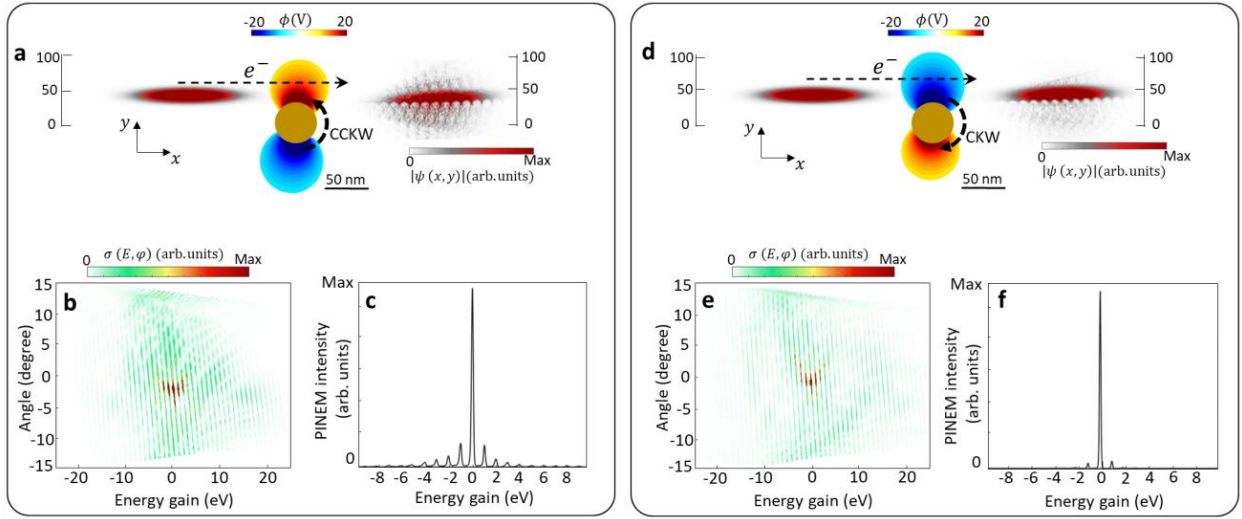

**Figure. S3. Impact of initial electron kinetic energy on coupling strength when the phase-matching condition is not satisfied**. Systems involving (a-c) CCKW and (d-f) CKW plasmon rotors. (a, d) Real-space representations of the electron wavepacket amplitude before and after propagating through the near-field region, excited by two perpendicularly polarized laser pulses with a $\pm\frac{\pi}{2}$ phase offset with the snapshots of the spatial profile of the scalar potentials depicted in the figures as well. (b, e) Angle-resolved inelastic scattering cross section of the probability amplitude after the interactions. (c, f) Energy-gain spectra obtained by integrating over the entire angular distribution. The electron has an initial kinetic energy of 100 eV. The laser pulses have an 800 nm carrier wavelength and a peak electric-field amplitude of $E_0 = 2$ GVm$^{-1}$.

The dependence of modulation efficiency on the kinetic energy of the electron reveals that, although the interaction time between a slow electron ($E_e = 100$ eV) and the near-field is prolonged, the initial interaction conditions do not satisfy the synchronicity criterion. Consequently, destructive interference between different electron pathways cancels out the manipulation, particularly along the longitudinal direction. Figures S. 3a and S. 3d illustrate the electron wavepacket before and after interaction with CCKW and CKW rotational near-fields, respectively, each with 800 nm wavelength. As shown in the coupling strength map, slow electrons with a kinetic energy of 100 eV propagating through the CCKW and CKW dipolar modes of a 25 nm gold nanorod exhibit a weak coupling strength. This results in minimal inelastic energy transfer, as evidenced by the appearance of only a few peaks in the PINEM spectra (Figures S3c and S3f). However, a broad angular deflection is observed in the inelastic-interaction cross-section maps (Figures S. 3 b and S3. e), resembling a part of an Ewald sphere. This deflection extends over a diffraction angle of approximately $\pm15°$ for both cases.

# Supplementary Note 4, 5: Dynamics of Diverged Electron Beam Manipulation by Time-Varying Near-Field in CCWS/ CWS Systems

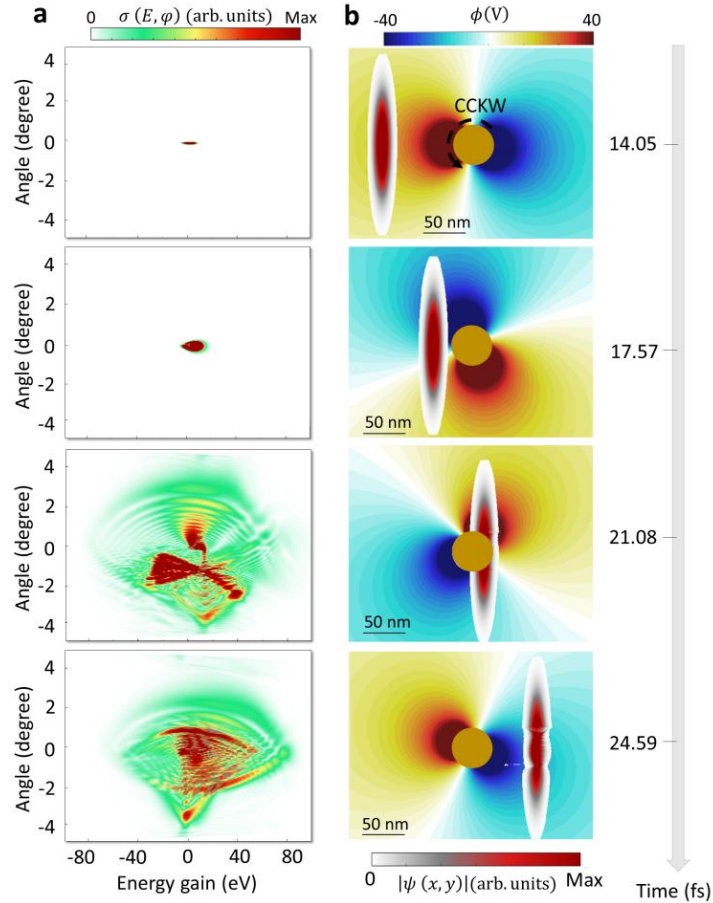

**Figure. S4. Dynamics of point-projected electron wavepacket modulation driven by CCKW plasmon rotors.** (a) Inelastic scattering cross-section during the interaction. (b) Evolution of the scalar potential and spatial representation of the electron wavepacket at specified time points. The gold nanorod is excited by two perpendicular laser pulses with a $+\frac{\pi}{2}$ phase lag. The laser pulses are characterized by a field amplitude of $E_0 = 2 \text{ GVm}^{-1}$, a wavelength of 800 nm, and a FWHM temporal broadening of 21 fs. The electron wavepacket is initialized with a kinetic energy of 1 keV and has longitudinal and transverse broadenings of 15 nm and 132 nm.

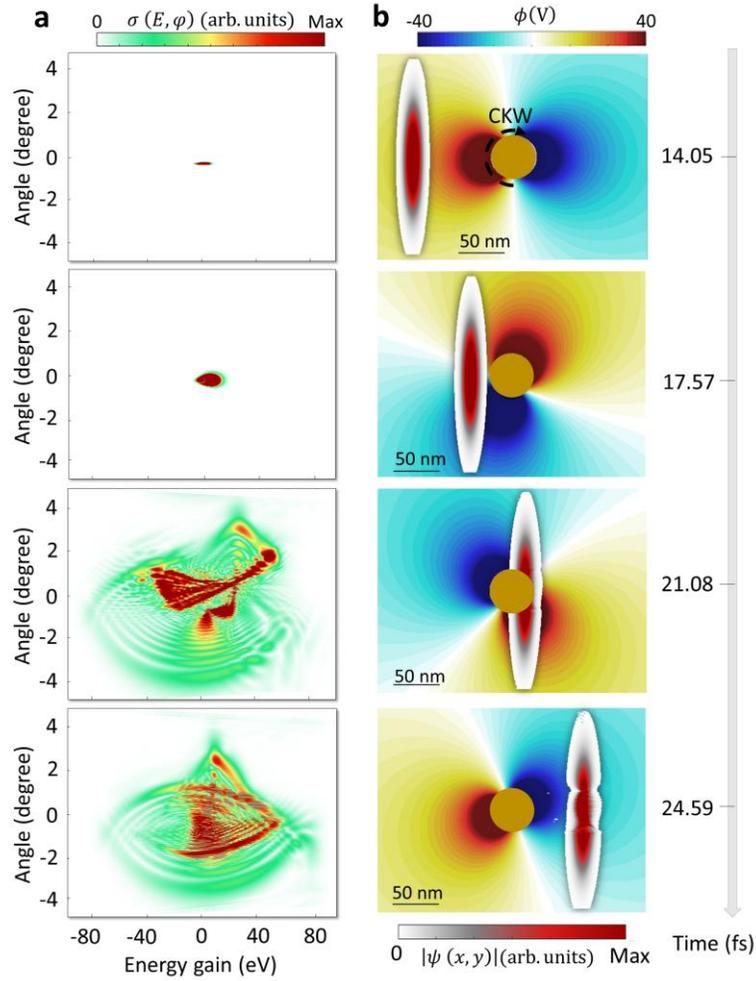

**Figure. S5. Dynamics of point-projected electron wavepacket modulation driven by CKW plasmon rotors.** (a) Inelastic scattering cross-section during the interaction. (b) Evolution of the scalar potential and spatial representation of the electron wavepacket at specified time points. The gold nanorod is excited by two perpendicular laser pulses with a $+\frac{\pi}{2}$ phase lag. The laser pulses are characterized by a field amplitude of $E_0 = 2$ GVm$^{-1}$, a wavelength of 800 nm, and a FWHM temporal broadening of 21 fs. The electron wavepacket is initialized with a kinetic energy of 1 keV and has longitudinal and transverse broadenings of 15 nm and 132 nm.

The spatiotemporal behavior of the electron wavefunction during its interaction with CCKW and CKW plasmonic near-fields are illustrated in Figures S4 and S5. The (a) columns show the inelastic scattering cross-section during the interaction, while the (b) columns depict the evolution of the scalar potential and the spatial representation of the electron wavepacket at specific times. As shown in Figure S4, an electron traveling along the *x*-axis retains its initial energy and momentum in free space. Upon entering the interaction region, the electron encounters four complete circulations of the near-field. For point-projection electrons, where the temporal coherence is orders of magnitude smaller than the wavelength of the oscillating field, the interaction time is short due to the narrow electron beam passing rapidly through the near-field region. When an electron enters the near-field zone of a CCKW gold nanorod, it experiences a rotational Lorentz force. This interaction induces a circular wiggling motion in the electron wavepacket, causing it to ascend the momentum ladder and populate higher states until $t = 17.57\ fs$. During the time interval $t = 17.57$ to 21.8 fs, as the electron reaches the center of the interaction zone, interference occurs between the upper and lower parts of the wavepacket. These parts experience different phases of the time-varying field with a fixed-phase difference. Hence this interference reshapes the electron modulation, where the two wavepacket parts mediate energy transfer and contribute to shaping the electron recoil. At $t = 24.59\ fs$, the Lorentz forces push the electron upward, forming a momentum representation resembling an eye-like pattern with a triangular structure centered at a positive diffraction angle. A similar process occurs when an electron interacts with a CKW rotational field, as shown in Supplementary Figure. S5. In this case, the field circulation is opposite to that of the CCKW scenario. Consequently, the electron is deflected downward in real space, and the momentum representation shows a summation of transverse ladder states centered around a negative diffraction angle.

## Supplementary Note 6: Calculation of Electron-Photon Interaction with Quasi-Static Approximation

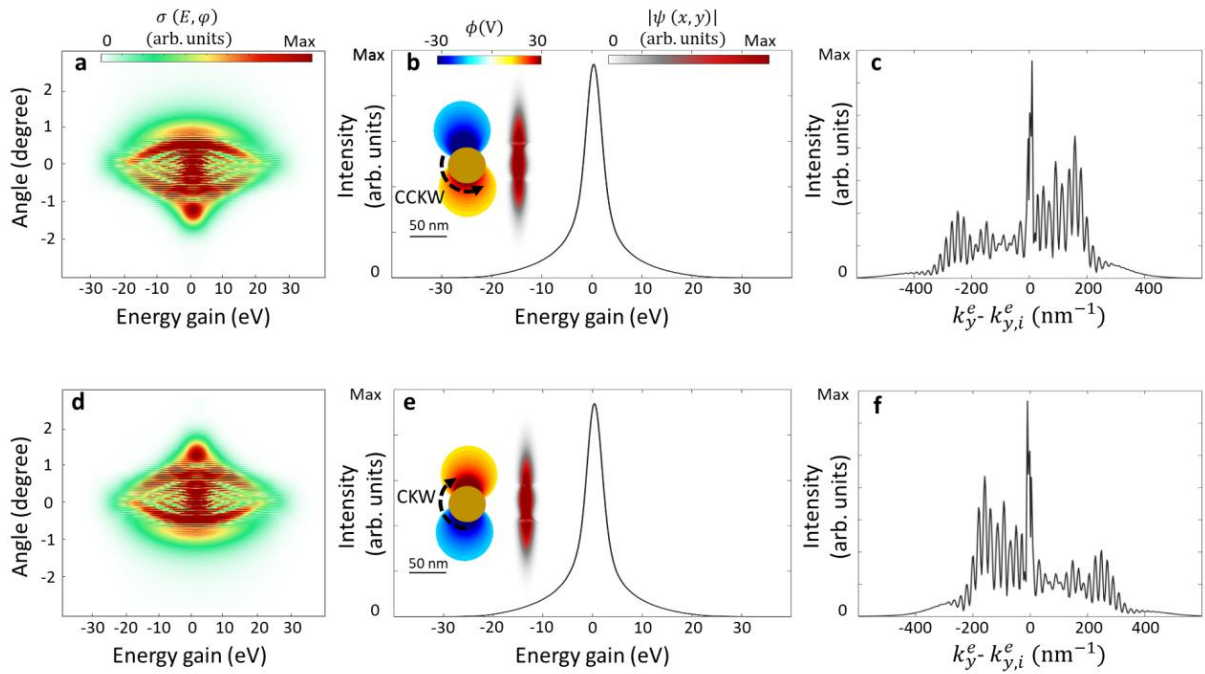

**Figure. S6. Electron modulation induced by plasmonic rotors analyzed with quasi-static approximations**. The electron wavepacket is shaped due to the interaction with (a-c) CCKW and (d-f) CKW induced plasmonic dipoles. (a) and (d) depict the inelastic scattering cross-section of the electron wavepacket. (b, e) PINEM spectrum, real-space distribution of the electron wavepacket, and a snapshot of the induced plasmonic near-field. (c, f) Transverse recoil of the electron beam integrated over the full energy range. The electron beam is characterized by a kinetic energy of 1 keV, with FWHM longitudinal and transverse broadenings of 15 nm and 132 nm, respectively. The laser pulses have both a central wavelength of 800 nm, and a peak field amplitude of 2 $GVm^{-1}$.

To better understand the role of each component of the Hamiltonian in electron-photon interactions, the quasi-static approximation is applied to simplify the analysis. As previously demonstrated, for sufficiently slow electrons, the scalar potential term dominates the interaction [13]. Therefore, Hergert, et al. have used the quasi-static approximation to model the interactions between slow-electrons and localized plasmons. Therefore, in this study, here, we consider only the classical scalar potential $\varphi(x, y, t)$, in the Hamiltonian[62]:

$$\widehat{H} = \frac{\hat{p}^2}{2m_0} - e\varphi(x, y, t), \tag{S1}$$

We modeled the propagation of a single-electron wavepacket with the initial conditions shown in Figure 3 of the manuscript and solved the time-dependent Schrödinger equation in two dimensions:

$$i\hbar \frac{d}{dt}\psi(x, y, t) = \widehat{H}\psi(x, y, t), \tag{S2}$$

Where $\hat{p} = -i\hbar \nabla$, and the time propagator in the simulations is approximated using a second-order differencing scheme.

To account for clockwise and counterclockwise rotational near-fields in the numerical solutions of this equation, we modeled the near-field potential as the summation of two orthogonal, linearly polarized monochromatic dipoles with a $\pm \frac{\pi}{2}$ phase difference. When linearly polarized lasers with spatially homogeneous amplitude $E_0$ optically excite a gold nanorod, they induce a local optical near-field with the total potential [62]:

$$\varphi_{Tot}(x, y, t) = \varphi_x(x, y, t) + \varphi_y(x, y, t), \tag{S3}$$

where each potential is driven by:

$$\varphi_x = \varphi_{x,0} \cos(\omega_{ph} t + \phi), \tag{S4}$$

$$\varphi_y = \varphi_{y,0} \cos(\omega_{ph} t + \phi \pm \frac{\pi}{2}),$$

Here, $\phi$ is the phase retardation relative to the incident laser field, induced by the complex dielectric function $\varepsilon$, and $\varphi_{x,0}$, $\varphi_{y,0}$ are the initial potentials for *x*-polarized and *y*-polarized excitations, respectively, recast as

$$\varphi_{x,0} = \begin{cases} E_0 x \left|\frac{\varepsilon - 1}{\varepsilon + 1}\right|, & r < R \\ E_0 x \frac{R^2}{r^2} \cdot \left|\frac{\varepsilon - 1}{\varepsilon + 1}\right|, & r \geq R \end{cases} \tag{S5}$$

$$\varphi_{y,0} = \begin{cases} E_0 y \left|\frac{\varepsilon - 1}{\varepsilon + 1}\right|, & r < R \\ E_0 y \frac{R^2}{r^2} \cdot \left|\frac{\varepsilon - 1}{\varepsilon + 1}\right|, & r \geq R \end{cases} \tag{S6}$$

where $R = 25$ nm, is the radius of the gold nanorod, and the laser field amplitude and wavelength are 2 GVm$^{-1}$, and 800 nm respectively.

Figure S6 illustrates the modulation of the point-projected electron after interaction with counterclockwise (Figures S6 a, b, c) and clockwise (Figures S6 d, e, f) near-fields, calculated using this approximation. Comparing these results with data from first-principles calculations (Figure 3 in the main text) shows that this approximation effectively captures the overall response and behavior of the system along both elastic and inelastic channels (e.g., inelastic scattering cross-section maps in Figures S6 a, d). However, the intensity and broadening of both energy transfer (Figure S6 b, e) and angular diffraction (Figure S6 c, f) are approximately half of those obtained from first-principles studies. This indicates that while the quasi-static approximation provides a good representation for near-field induced slow electron modulation, it fails to fully capture all aspects of the interaction, especially for higher-order effects and broader energy distributions, mainly due to neglecting the vector potential.